\documentclass[sigconf]{acmart}

\usepackage{booktabs} 
\usepackage{colortbl} 
\usepackage{multirow}
\usepackage{url}
\usepackage[show]{chato-notes}
\setcopyright{acmcopyright}
\usepackage{appendix}

\acmDOI{https://doi.org/10.1145/3605098.3636044}

\acmISBN{979-8-4007-0243-3/24/04}

\acmConference[SAC'24]{ACM SAC Conference}{April 8–April 12, 2024}{Avila, Spain}
\acmYear{2024}
\copyrightyear{2024}

\acmArticle{4}
\acmPrice{15.00}

\begin{document}
\title{Evaluating Trustworthiness of Online News Publishers\\ via Article Classification}
  
\renewcommand{\shorttitle}{Evaluating Trustworthiness of Online News Publishers}

\author{John~Bianchi}
\authornote{Corresponding Author}
\orcid{xyz}
\affiliation{%
\institution{IMT School for Advanced Studies Lucca}
\country{Italy}
\postcode{55100}  
}
\email{john.bianchi@imtlucca.it}

\author{Manuel~Pratelli}
\orcid{0000-0002-9978-791X}
\affiliation{%
\institution{IMT School for Advanced Studies Lucca\\ IIT-CNR}
   \country{Italy}
   \postcode{55100}  
 }
\email{manuel.pratelli@imtlucca.it}

 \author{Marinella~Petrocchi}
 \affiliation{%
   \institution{IIT-CNR\\ IMT School for Advanced Studies Lucca}
   \country{Italy}}
 \email{marinella.petrocchi@iit.cnr.it}

 \author{Fabio~Pinelli}
 \affiliation{%
   \institution{IMT School for Advanced Studies Lucca}
   \country{Italy}
 }

 \email{fabio.pinelli@imtlucca.it}

\renewcommand{\shortauthors}{J. Bianchi et al.}

\begin{abstract}
 The proliferation of low-quality online information in today's era has underscored the need for robust and automatic mechanisms to evaluate the trustworthiness of online news publishers. In this paper, we analyse the trustworthiness of online news media outlets by leveraging a dataset of 4033 news stories from 40 different sources. We aim to infer the trustworthiness level of the source based on the classification of individual articles' content. The trust labels are obtained from NewsGuard, a journalistic organization that evaluates news sources using well-established editorial and publishing criteria. The results indicate that the classification model is highly effective in classifying the trustworthiness levels of the news articles. This research has practical applications in alerting readers to potentially untrustworthy news sources, assisting journalistic organizations in evaluating new or unfamiliar media outlets and supporting the selection of articles for their trustworthiness assessment.
\end{abstract}

%
%



\maketitle

\section{Introduction}
\label{sec:intro}
{\it Disintermediation}, or the phenomenon of reducing intermediate flows, is a term coined back in 1983, when author Paul Hawken called by this name the set of processes by which consumers could directly manage financial investments in securities, rather than leaving their money in savings accounts~\cite{Hawken83}.


Over time, many industries have experienced disintermediation. In tourism, the Internet provides users with access to a wealth of information, allowing them to seamlessly assemble various tourism services and create unique travel experiences on their own. Similarly, the growing trend of self-publishing places increasing responsibility on authors to oversee the entire process of producing and distributing their work~\cite{Forbes2022}). Journalism has also experienced changes that reflect the evolving landscape of direct access to information and news dissemination. These shifts indicate a broader societal trend toward greater autonomy and control in various industries.
Particularly with regard to journalism, the emergence of new web technologies and social networks has diminished the essential role of traditional journalists as the prevalence of participatory journalism facilitated by blogs and social networks continues to grow~\cite{Jemielniak2020,Bowman2003}. In this regard, a recent UNESCO report\footnote{\url{https://news.un.org/en/story/2022/03/1113702}. All of the URLs in this document were last accessed on December 22, 2023.} on the existential threat posed by social media to traditional news claims that online `news outlets often struggle to get the clicks from readers that determine advertising revenue' and job cuts in journalism have resulted in a noticeable void in the information landscape. 

The erosion of the mainstream journalism system in recent years, coupled with challenges such as understaffing and the pressure to publish attention-grabbing news to re-engage readers, has raised concerns about the quality of information provided by online media. Various journalism organizations and indices, including NewsGuard\footnote{NewsGuard: https://www.newsguardtech.com}, the MediaBias Fact Check\footnote{MediaBias Fact Check: https://mediabiasfactcheck.com/}, the Iffy index of unreliable sources\footnote{Iffy Index: https://iffy.news/index/}, the Global Disinformation Index\footnote{The Global Disinformation Index: https://www.disinformationindex.org/}, the Ad Fontes Media\footnote{Ad Fontes Media: https://adfontesmedia.com/} that conduct studies on the transparency and trustworthiness of online news sources, including their tendency to produce propagandistic and/or politically biased content. 

Although different organizations use different criteria to determine the trustworthiness of an online media outlet, recent work has found excellent convergence in the labels each assigns to individual media outlets, confirming the degree of trustworthiness of the judgments~\cite{PNASPennycook2023,DBLP:conf/icwsm/PratelliP22}. 

Unfortunately, the process of evaluating each news outlet is very cumbersome, especially in terms of time. For example, the procedure followed by the Global Disinformation Index is to select annotators who are experts in the online information system of a particular country. 
After training, the annotators select a group of newspapers that accurately reflect the country's information landscape. They then manually analyze these newspapers to find information on aspects such as ownership and funding sources.
This is followed by a manual content analysis of a sample of articles per media outlet to check for unreliable, sensational and/or propagandistic content. 
The GDI then processes the results of the study, which are summarized in a score between 0 and 100 that indicates the risk that the media outlet is misinforming its readers\footnote{The second and third authors are familiar with the procedure, having participated as annotators in the GDI country study on the Italian online media market \cite{Petrocchi22GDI}.}.

In this paper, we aim to label the trustworthiness level of a news source from the classification of the news itself. 
Operationally, we start with a dataset of $4033$ news stories from $40$ online news outlets, which we have collected and to which we have attached labels regarding both the main topic and the trustworthiness score of the news outlet.  The labels are collected by NewsGuard, which is licensed to the authors of this paper.
Through qualified journalists, NewsGuard rates all news sources, which account for 95 percent of online engagement\footnote{https://www.newsguardtech.com/solutions/newsguard/}.
Each site is analyzed according to nine accepted journalistic criteria. Based on these nine criteria, the site receives a trustworthiness score from 0 to 100. The trustworthiness levels are 5, ranging from `high credibility,' the best rating, to `proceed with extreme caution,' indicating a site with a very low level of transparency and credibility. 

On the one hand, tagging the articles in our dataset with NewsGuard labels relieves us of the tedious task of annotating the data and gives us a solid ground truth based on the work of specialized journalists. 
On the other hand, before moving on to the main goal of the work, which is to derive the level of trustworthiness of the news source from the analysis of individual articles, we will test the goodness of the dataset by classifying articles by topic and making sure that the predicted topic matches the topic assigned in the label.

We combine 3 standard topics, i.e., {\it Sports}, {\it Political News} and {\it Health}, with an escalating one, in the age of the internet and pandemics, vaccines and wars, namely {\it Conspiracy Theories}~\cite{Conspiracy2021}.

\paragraph{Results}
Our models have proven to be highly effective in both classification tasks. The results are summarized as follows:

\begin{itemize}

    \item \textbf{Trustworthiness Detection Task:} This is a multi-class classification task at the article level. Our model successfully predicts the level of trustworthiness of news sources based on the article text alone. Specifically, we obtain an 
    average F1-macro of $0.843$ and an average F1-micro of $0.882$ for this task (see section \ref{sec:trustDetection}).
    
    \item \textbf{Topic Detection Task:} This is a multi-class classification task at the article level. Our model achieves an 
    average F1-macro of $0.925$ and an average F1-micro of $0.929$. 
    This gives us an additional level of confidence in the original NewsGuard labeling of the data set.  Wrong predictions arise when distinguishing between \textit{`Conspiracy'} and (\textit{`Health"} or \textit{`Political'}), leading to some misclassifications (see Section \ref{sec:topdec}).

    
\end{itemize}

\paragraph{Applications}

The ability to predict a publisher's level of trustworthiness from the classification of individual articles suggests at least three possible applications, one to assist the user, the others to assist journalistic organizations (e.g., NewsGuard and GDI):

\begin{enumerate}
\item 
at the user level, the classifier can be used as a tool to alert the reader by displaying a meaningful visual signal, such as the classic red flag, while the reader is viewing an article from an unfamiliar news outlet. The red flag could say something like `the article you are reading is similar to those produced by untrustworthy newspapers. Supplement your reading with other readings of articles produced by trustworthy newspapers'. 
\item at the organizational level, let the reader assume that the media outlet is new or completely unknown to the evaluator (which is very common these days given the constant proliferation of alternative online media outlets~\cite{Alternative2022report}). An initial idea of its level of trustworthiness can be obtained by collecting a number of articles and applying the trustworthiness ranking model to them. At a later date, if the evaluator deems it necessary, a more comprehensive investigation can be conducted using traditional journalistic analysis.
\item still at the organizational level, the selection of articles to analyze to assess the publisher's trustworthiness are typically the most shared articles on social media and/or articles containing a set of representative keywords.\footnote{\url{https://www.disinformationindex.org/country-studies/2023-06-08-disinformation-risk-assessment-the-online-news-market-in-thailand/}}. Unfortunately, this method may not be sufficient to select a sample of articles that is truly representative of the publisher. For example, if we relied on the most shared articles on social media, we might select a sample consisting only of straight news stories (e.g., traffic accidents, robberies, etc.).
Therefore, we argue that the models presented in this paper can be used to process a selection of articles from the target media for a more balanced sample that can be manually analyzed. This approach ensures a balanced assessment in terms of both trustworthiness levels and topics.
\end{enumerate}

\section{Problem definition}\label{sec:problemDef}

In this section, we formalize the main problem addressed in this paper, called \textit{Trustworthiness Level} Detection, and present the performance metrics used to evaluate the resulting models. The \textit{Topic Detection} task is also formalized, since we use the results of this classifier to experimentally evaluate the quality of the labeling procedure obtained from NewsGuard.

Let be \(A\) a set of articles. This set represents a collection of articles characterized by their textual content. Formally, \(A = \{a_1, a_2, \ldots, a_n\}\), where \(n\) is the total number of articles. Each article \(a\) has attributes \(text_a\) representing the textual content of the article and \(newspaper_a\), the newspaper from which the article originates.

The set of Newspapers \(N\) comprises various newspapers, each associated with a specific level of trustworthiness. Formally, \(N = \{n_1, n_2, \ldots, n_m\}\), where \(m\) is the total number of newspapers. Each newspaper \(n_i\) has \(trust_{n_i}\) that represents the level of trustworthiness associated with it.


Notice that different levels of trustworthiness associated with newspapers belong to the set \(L\). Formally, \(L = \{l_1, l_2, \ldots, l_p\}\), where \(p\) is the total number of trustworthiness levels. The level of trustworthiness is typically a continuous value, but we prefer to aggregate values in bins that correspond to our levels of trustworthiness \(L\). We provide more details about the bins in Section~\ref{sec:dataset}. 

Notice that each article \(a_i\) is associated with one and only one level of trustworthiness (\(trust_{a_i}\)) from the set of trustworthiness levels (\(L\)) inherited by the newspaper it originated from.

Therefore, given the sets \(A\), \(N\), and \(L\), and the constraints mentioned above, we formalize the problem as follows: we aim to develop a text classifier \(C_{trust}\) that associates the level of trustworthiness \(trust_a \in L\) for each article \(a\) based on its textual content \(text_a\).
The Classifier \(C_{trust}\) is a machine learning model that addresses the \textit{Trustworthiness Level} Detection task, and it relies on the NewsGuard annotation process described in Section~\ref{sec:intro}.
Given this, we can reasonably assume that the level of trustworthiness associated with each newspaper is reliable, a notion further supported by considering the primary topic associated with each newspaper.


To this end, we define the \textit{Topic Detection} task as follows. Each newspaper \(n_i\) has an additional attribute associated with it: \(topic_{n_i}\) that indicates the topic it primarily covers. 
This information is gathered by NewsGuard. 
The set \(T\) defines the possible categories or topics into which articles can be classified. Formally, \(T = \{t_1, t_2, \ldots, t_k\}\), where \(k\) is the total number of topics.
As for the previous classification task, each article \(a_i\) is associated with one and only one Topic (\(topic_{a_i}\)) from the set of Topics (\(T\)) inherited by the newspaper it originated from. Thus, given the sets \(A\), \(N\), and \(T\), and the constraints mentioned above, we formalize the problem as follows: we aim to develop a text classifier \(C_{topic}\) that associates a \(topic_a \in T\) for each article \(a\) based on its textual content \(text_a\). The Classifier \(C_{topic}\) is a machine learning model for \textit{Topic Detection} task. As in~\cite{Semeval23Task3}, which addressed a similar challenge to ours, the performance of the classifiers is evaluated using the F1 Micro and F1 Macro metrics, defined as follows: 

\begin{equation}\label{eq:F1Micro}
    \text{F1 Micro} = \frac{2 \cdot \text{Precision} \cdot \text{Recall}}{\text{Precision} + \text{Recall}} 
\end{equation}

\begin{equation}\label{eq:F1Macro}
    \text{F1 Macro} = \frac{1}{|M|} \sum_{i=1}^{|M|} \frac{2 \cdot \text{Precision}_i \cdot \text{Recall}_i}{\text{Precision}_i + \text{Recall}_i}
\end{equation}

Using these two metrics allows us to apply a consistent measurement approach, as established by \cite{Semeval23Task3}, and thereby illuminate two different facets of classification performance. The F1 Micro (\ref{eq:F1Micro}) considers all predictions comprehensively, while the F1 Macro (\ref{eq:F1Macro}) treats each class independently by calculating a weighted average of the F1 scores for each class.
In Equation \ref{eq:F1Macro}, we denote a general set of classes \(M\). In our tasks, the classes are defined by the set \(L\) representing levels of trustworthiness for the \textit{Trustworthiness Level} Detection and the set \(T\) representing topics for the \textit{Topic} Detection.

\section{Dataset}\label{sec:dataset}

This section outlines the steps leading to the final dataset of articles used in our experiments: (i) selecting a representative list of online media, i.e., the \(N\) set of newspapers introduced above, (ii) retrieving the textual content of the articles published by the selected sources, i.e., the \(A\) set of articles, and (iii) data cleaning.



\subsection{Online Media Outlets Selection}\label{sec:SourceSelection}


The goal of this selection process is to create a fair and representative list of online media that accurately reflects real-world conditions and allows for effective testing of our models.
To achieve this goal, we started with the NewsGuard dataset of tagged online news sources, which is available to authors under the NewsGuard license. This dataset contains news sources that have been rigorously evaluated by expert journalists. Notably, the news sources within this dataset have significant impact, collectively contributing to 95\% of online engagement. 
Our focus within this dataset is on specific source-level attributes, namely, `topics' and `trustworthiness scores'. We have identified four key topics for the analysis: {\it Political news or commentary}, {\it Conspiracy theories or hoaxes}, {\it Sports and athletics}, and {\it Health or medical information}, i.e., the set \(T\) introduced in Section \ref{sec:problemDef}. Starting with the original dataset, we first selected the sources associated with one and only one topic.

Now we want to obtain a list of news sources, 10 per topic, that preserves the distribution of NewsGuard's trust scores for each selected topic. In other words, we want to obtain a set of news sources that represent the true distribution of trust scores in the original dataset. We use a stratified sampling approach based on the trustworthiness levels, i.e., the set \(L\) of Section~\ref{sec:problemDef}, defined by NewsGuard (see Table \ref{tab:newsguardBins}).

For example, let the reader consider the online sources related to the topic {\it Sports and athletics}. The distribution of trustworthiness scores for these sources is as follows: 30\% of them have a trustworthiness score of 100; 50\% fall within the range 75-99, and 20\% fall within the range 60-74. So, to keep the original distribution, we randomly take 3 sources with $l$=100, 5 sources with $l$=75-99, and 2 sources with $l$=60-74.

Assuming that the original NewsGuard dataset accurately represents the actual online media landscape, our stratified sampling approach allows us to obtain 40 sources that closely match and mirror the original NewsGuard dataset in terms of the distribution of trustworthiness across the four specified topics. 
Finally, we exclude sources with a paywall and those that are not in English. 
This represents our  set \(N\) of newspapers.

\begin{table}[!h]
    \centering
    \caption{\label{tab:newsguardBins}Newsguard Trustworthiness Levels 
    }
    \begin{tabular}{ll}
        \hline
        \textbf{Score} & \textbf{Description} \\ \hline
        100 & High Credibility \\ \hline
        75-99 & Generally Credible \\ \hline
        60-74 & Credible with Exceptions \\ \hline
        40-59 & Proceed with Caution \\ \hline
        0-39 & Proceed with Maximum Caution \\ \hline
    \end{tabular}
\end{table}

\subsection{Articles collection}\label{sec:dataCollection}
The next stage is to collect a sample of articles for each selected source. Specifically, we collect the textual content of the most recently published articles on the sources' websites. The goal is to generate the set \(A\) of articles presented in \ref{sec:problemDef}.

To accomplish this task, we follow a multi-step process. We compile a list of URLs corresponding to the published articles. From the homepage addresses of the media outlets reported in the NewsGuard dataset, we manually identify the web page that contains the list of published articles (the so-called `news history' page). 

\begin{figure}[h!]
    \centering
    \includegraphics[width=.9\columnwidth]{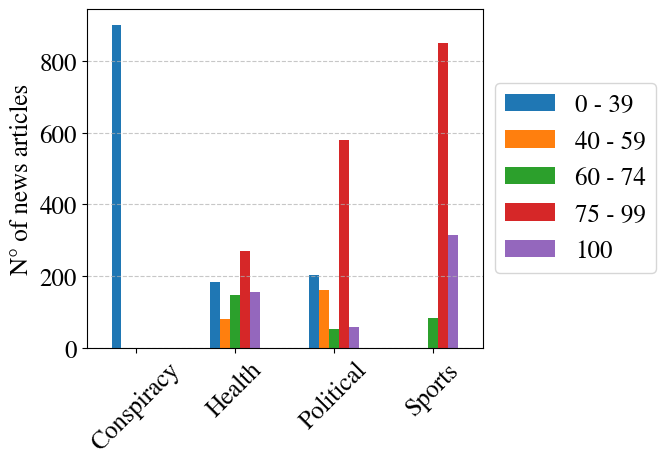}
    \includegraphics[width=.9\columnwidth]{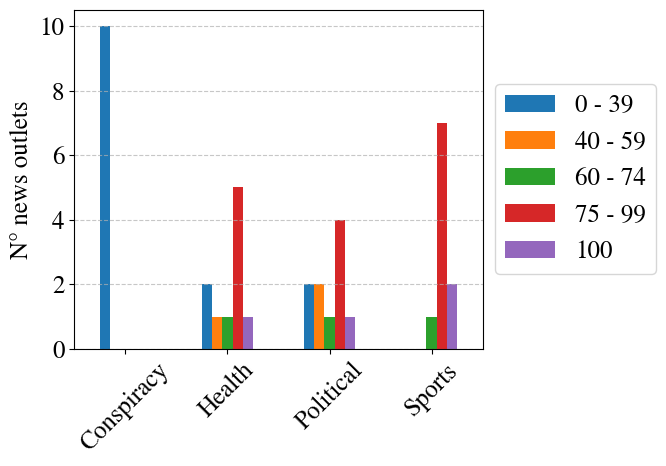}
    \caption{Number of articles (top) and news outlets (bottom) per trustworthiness level, broken down by topic.}
    \label{fig:trustRangePlot}
\end{figure}

We use Selenium library\footnote{\url{https://www.selenium.dev/}} to develop a script that automatically scrolls through these web pages to collect the necessary URLs. For example, if we look at the website of \textit{The Sun}\footnote{\url{https://www.the-sun.com/}}, the news history page is available at:
\textit{https://www.the-sun.com/news/us-news/\textbf{page/1/}}. The developed application systematically jumps to the next pages by increasing the number in the URL. In the example, it becomes \textit{https://www.the-sun.com/news/us-news/\textbf{page/2/}}, and for each of these pages, it collects all the URLs related to individual news articles. Once we have the URL lists for each source's articles, we retrieve the HTML content of each article page referenced in the URL list. Specifically, we utilize the GNU Wget\footnote{\url{https://www.gnu.org/software/wget/}} command to retrieve and save the complete HTML content of these pages in WARC-format archives. This ensures a readily accessible offline copy of the HTML pages, facilitating subsequent textual content extraction. The final step involves the creation of custom text extractors tailored to each news source. These extractors utilize XPATH-based information to effectively extract the embedded text from each article's web page. We do not consider news articles with less than 200 words; we acquire at least 40 articles per news outlet, with a maximum of 294. Thus, out of 5006 articles, we collected the texts of 4033 of them. The resulting dataset \(A\), as depicted in Figure \ref{fig:trustRangePlot} (top and bottom) and Table \ref{tab:articleByTopic}, comprises articles extracted from the selected media outlets \(N\) during the period spanning from May 4th to May 15th, 2023. The figures provide an overview of the distribution of $l$ per topic for the 40 news outlets. The {\it Conspiracy} topic is entirely dominated by articles falling within the lowest score range. For \textit{Health or medical information} and \textit{Political news or commentary}, the data show a more evenly distributed pattern, with a prevalence in the $l$=75-99 score range for the number of sources in the latter topic. \textit{Sports and athletics}, which includes the largest number of articles, 1246, presents a high concentration of elements in the $l$=75-99 interval.

\begin{table}
    \centering
    \caption{Number of Articles per trustworthiness level, broken down by topic \label{tab:articleByTopic}}
    \begin{tabular}{lcccc}
        \toprule
        \textbf{Levels} & \textbf{Political} & \textbf{Conspiracy} & \textbf{Sports} & \textbf{Health} \\
        \midrule
        0 - 39 & 204 & 900 & 0 & 183 \\
        40 - 59 & 162 & 0 & 0 & 79 \\
        60 - 74 & 51 & 0 & 83 & 146 \\
        75 - 99 & 579 & 0 & 849 & 269 \\
        100 & 59 & 0 & 314 & 155 \\
        \midrule
        \textbf{Total} & \textbf{1055} & \textbf{900} & \textbf{1246} & \textbf{832} \\
        \bottomrule
    \end{tabular}
\end{table}

 \subsection{Data cleaning}

While our XPATH-based extraction method (see Subsection \ref{sec:dataCollection}) provides flexibility by adapting to how each source organizes article content on its site, it often results in the inclusion of extraneous text fragments. These can be divided into two groups: (i) repetitive phrases (e.g., thank you messages, signatures, slogans) and (ii) other miscellaneous linguistic elements (e.g. dates).

We therefore clean the data by applying both \textit{Spacy}\footnote{\url{https://spacy.io/}},  specifically by utilizing the "en\_core\_web\_lg" model, and manual verification.
%
Statistics about the final dataset are in Table~\ref{tab:articleByTopic}. This dataset represents the set \(A\) of articles of the problem definition (see Section~\ref{sec:problemDef}).

\section{Results and discussion}\label{sec:results}
We present the experimental setup and results for the tasks defined in Section~\ref{sec:problemDef}, specifically, \textit{topic} and \textit{trustworthiness} detection. We use BERT~\cite{devlin2018bert}, the well-known state-of-the-art pre-trained language model.
We use the \textit{transformers}\footnote{\url{https://github.com/huggingface/transformers}} library in Python to deploy and fine-tune BERT, as well as to compute performance metrics. In particular, we adopt \textit{BertForSequenceClassification} because it combines the capabilities of a highly trained language model with the adaptability to address specific tasks. For the validation, we adopt the 10-fold stratified cross-validation implemented by \textit{scikit-learn}\footnote{\url{https://scikit-learn.org/stable/modules/generated/sklearn.model_selection.StratifiedKFold.html}},  thus using 10\% of the dataset as test and 90\% as training in each step. This choice guarantees an even distribution of the target class across each fold, which ensures a more robust evaluation.

\subsection{Topic detection}\label{sec:topdec}
Before we continue with the multi-class
classification task, we show the results of the topic detection task. We remind the reader that the execution of this task and the evaluation of its results should not be considered as the main result of this article. The task is valuable for understanding the quality of the dataset annotations, which we did not provide ourselves.

\begin{figure}
    \centering
    \includegraphics[width=.7\columnwidth]{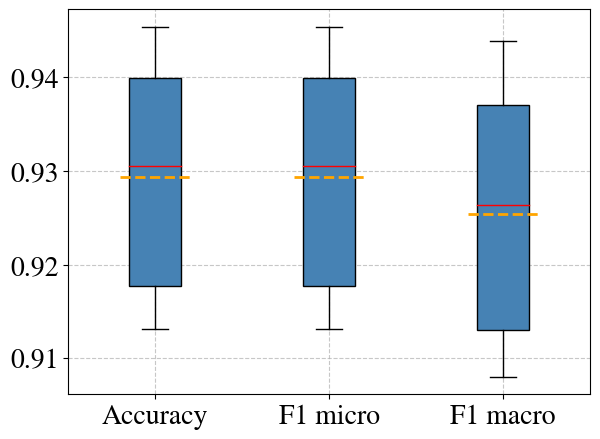}
    \caption{Evaluation results for topic detection. The yellow dotted line represents the average, while the red line represents the median.}
    \label{fig:smallTopicPerformance}
\end{figure}

Figure~\ref{fig:smallTopicPerformance} shows the evaluation results on the 10 stratified folds.
We achieve an average F1-macro of $0.925$ (min = $0.908$ and max = $0.943$) and an average F1-micro of $0.929$ (min = $0.913$ and max = $0.945$). For completeness, we also report the accuracy, precision and recall values in Figure \ref{fig:topicScores} in the appendix, which are quite high for each topic. 

\begin{figure}[h!]
    \centering
    \includegraphics[width=.8\columnwidth]{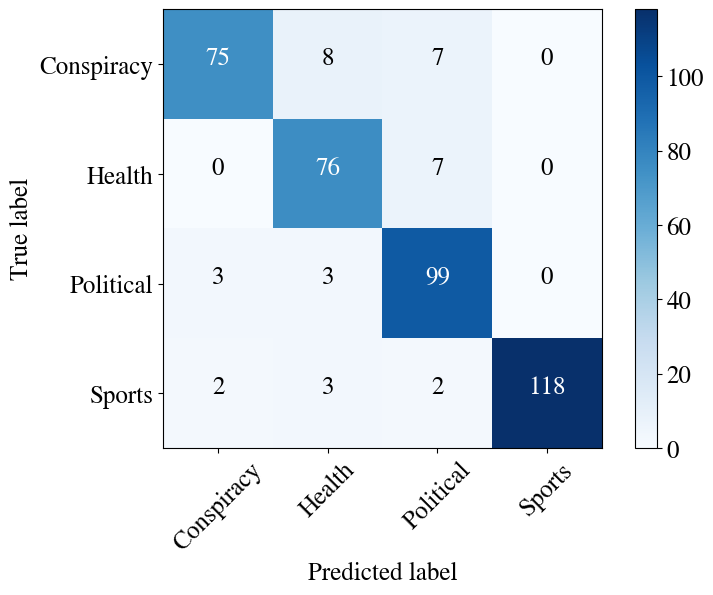}
    \includegraphics[width=.8\columnwidth]{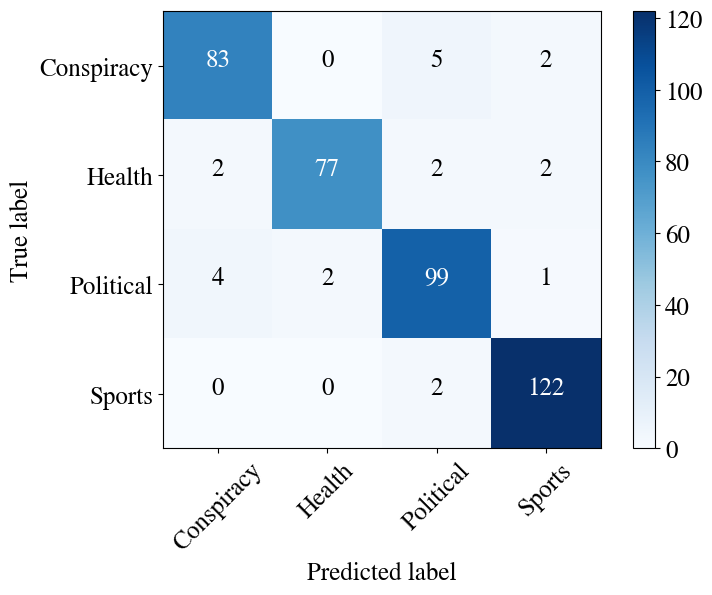}
    \caption{Topic: Confusion matrix for the fold with the lowest (top) and the highest (bottom) F1 macro}
    \label{fig:topicMatrix}
\end{figure}

Figure~\ref{fig:topicMatrix} (top) shows the confusion matrix for the fold with the worst F1 macro score. The errors made by the classifier are mainly due to the misclassification of {\it Political news or commentary} articles as \textit{Conspiracy theories or hoaxes}.
This result is not surprising, since 
the lines between legitimate political news and conspiracy theories can be blurred by information manipulation strategies~\cite{douglas2019understanding}. 
Also, some conspiracy articles are mistakenly categorized as related to health or medical information. This is not surprising since conspiracy theories often touch on topics related to public health, a phenomenon that has become more pronounced during and after the COVID-19 global pandemic~\cite{pummerer2021societal}. This level of misclassification is not observed when examining the best-performing fold (Figure \ref{fig:topicMatrix} bottom), where errors still exist, albeit to a lesser extent. Despite some inaccuracies, we argue that these results are satisfactory in that there is a very good match between articles and assigned topics.

\subsection{Trustworthiness Detection}\label{sec:trustDetection}
This section presents the results of detecting articles' trustworthiness level as defined in Section~\ref{sec:problemDef}. The experiments are performed on the final dataset described in Section~\ref{sec:dataset}, following the methods described at the beginning of this section. 

The primary goal of this task is to develop a classifier capable of assigning a level of trustworthiness ($\text{trust}_a$) to each article (\textit{a}) based on its textual content ($\text{text}_a$). These trustworthiness levels, which include five different categories identified and assigned by NewsGuard, provide a nuanced characterization of publisher-level trustworthiness (see Table \ref{tab:newsguardBins}). 
This is a multi-class classification
task at the article level.

\begin{figure}
    \centering
    \includegraphics[width=.7\columnwidth]{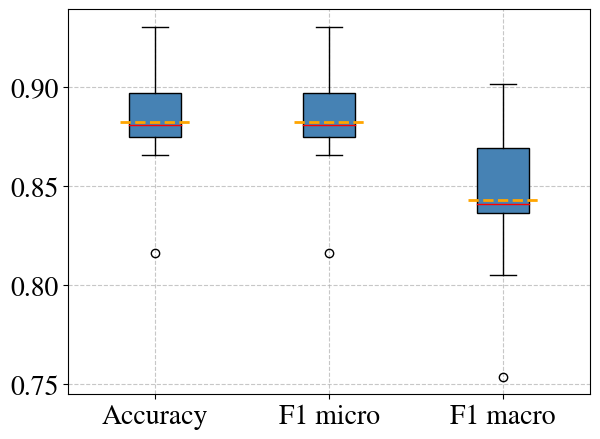}
    \caption{Evaluation results for trustworthiness level detection. The yellow dotted line represents the average, while the red line represents the median.}
    \label{fig:smallTrustPerformance}
\end{figure}

Our results, shown in Figure~\ref{fig:smallTrustPerformance}, demonstrate the strong capability of the model to accurately associate the article to one of the five trustworthiness levels. The model achieves an average F1-macro of $0.843$ (min = $0.753$ and max = $0.901$) and an average F1-micro of $0.882$ (min = $0.816$ and max = $0.930$).

\begin{figure}[h!]
    \centering
    \includegraphics[width=.8\columnwidth]{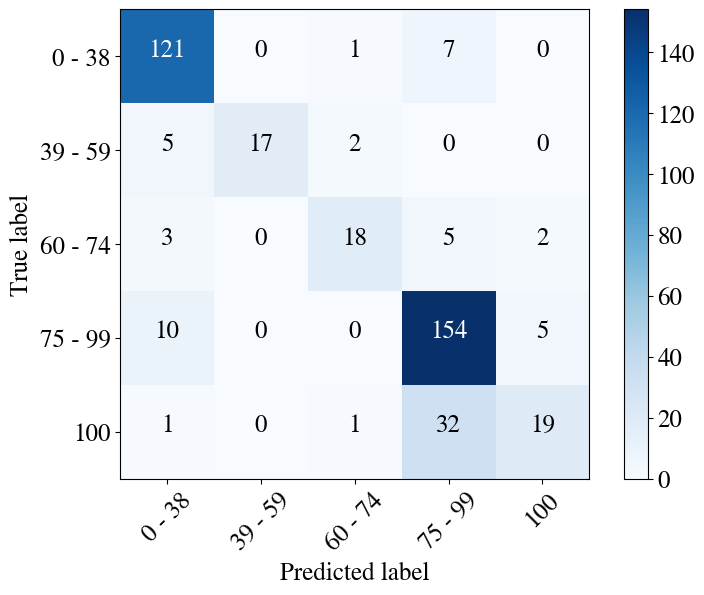}
    \includegraphics[width=.8\columnwidth]{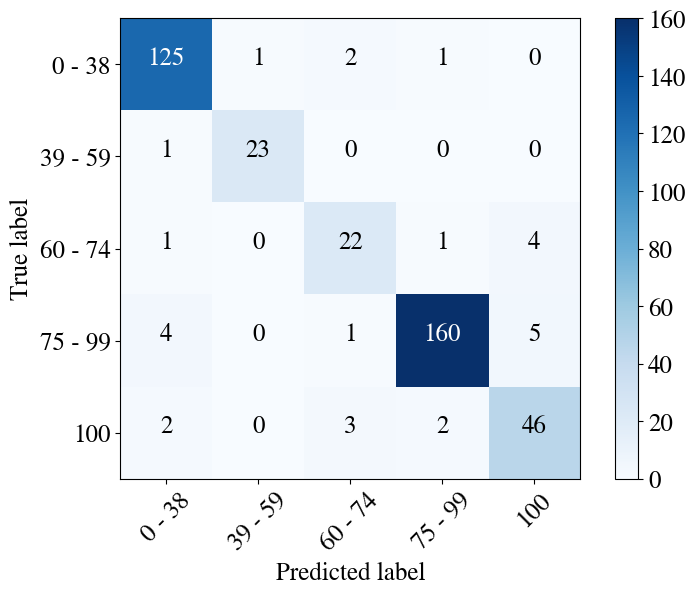}
    \caption{Trustworthiness level: Confusion matrix for the fold with the lowest (top) and highest (bottom) F1 macro}
    \label{fig:trustMatrix}
\end{figure}

We analyze the confusion matrices associated with the best and worst Macro F1 scores to gain deeper insight into the results. 
Figure~\ref{fig:trustMatrix} (top) shows the confusion matrix for the fold associated with the lowest Macro F1 score. Here we can observe two errors: 49 items assigned to neighboring classes and 25 items assigned to classes significantly different from the true ones. In the best case, as shown in Figure \ref{fig:trustMatrix} (bottom), the situation is characterized by lower values, with 17 items assigned to adjacent classes and 11 to distant classes.

The importance of the two types of errors can vary depending on how we want to use the model. 
As mentioned earlier, the trustworthiness levels represent a nuanced characterization of trustworthiness. There may also be situations where a coarser classification is desired. For example, we might consider redefining NewsGuard's thresholds to produce only two levels of trustworthiness: the $0-59$ and $60-100$ ranges to identify untrusted and trusted publishers. Redefining the thresholds to create coarser levels of trustworthiness can improve the performance of our model, thereby increasing its practical utility in real-world scenarios.

\section{Related Work}
Online news consumption is now vital for insights into societal and cultural trends. Automated analysis, particularly through machine/deep learning, proves crucial in categorizing web-based news. A recent survey~\cite{SAI2020} identified 51 studies (2000-2019) using supervised and unsupervised learning and generative models.


Early work was primarily aimed at testing the ability of traditional supervised machine learning tools, such as Support Vector Machines, to recognize news topics. The analysis was conducted on texts in English as well as in other languages, such as German~\cite{scharkow13}, and the features used were mostly n-grams or whole words extracted via the BoW method~\cite{10.5555/3110856}.

Among the works using `traditional' approaches, an interesting work is ~\cite{dai-etal-2018-fine}, where the authors aim not so much to identify the topic of the news item as to identify its structure. In publishing, a news story is defined, for example, as having an `Inverted Pyramid' structure when there is a so-called {\it lede} that introduces the story and a body section that, in an expository style, sets out other facts of the news. The paper considers both literal features and the writing style of the news to classify them according to their structure, e.g., the Inverted Pyramid, the Martini Glass, and the Kabob structures~\cite{heravi2022storytelling}. 

As we approach today, research focuses not only on news topics and structure recognition but also on detecting propagandistic, biased or untruthful content. This is the case with various challenges launched over the years, such as, for example, the Semantic Evaluation series, {\it Semeval}~\cite{semeval/MartinoBWPN20}, and CheckThat!@CLEF, which, for example, in the 2022 edition dealt with the infodemic phenomenon developed during the Covid-19 pandemic, launching tasks on recognizing tweets worth checking based on their veracity, and useful for fact checking others~\cite{10.1007/978-3-031-13643-6_29}. Task 3 of the 2023 edition of Semeval 2023~\cite{Semeval23Task3}, entitled `SemEval-2023 Task 3: Detecting the Category, the Framing,
and the Persuasion Techniques in Online News in a Multi-lingual Setup', extended the search for persuasion techniques in the news to include two additional types of categorization, {\it category} (namely, whether the text read represents opinion, objective news or satire) and {\it framing}\footnote{In communication research, `to frame' means `to select some aspects of a perceived reality and make them more salient in a communicating text'~\cite{Entman1993,liu-etal-2019-detecting,card-etal-2015-media}.}. Although it is not the focus of our paper, the topic detection task we performed in Section~\ref{sec:topdec} to assess the good match of topics to articles in our dataset can be considered, to some extent, a framing detection task. A conspiracy theory could be presented with a political, sports, or health framing. Conversely, news about politics, sports, and health might be written in a `conspiracy' style and be associated with conspiracies.
Nevertheless, SemEval-2023 Task 3 is by no means comparable with our work, the former being much more complex, with 14 possible frames, with articles associable with more than one frame, and multilingual. 

\paragraph{On the Evaluation of News Publisher's Trustworthiness }
The publisher's trustworthiness is a feature most often used to discern between true and fake news~\cite{bazmi2023multi,doi:10.1080/07421222.2019.1628921}. Assessing this value is crucial for studying the spread of misinformation online~\cite{lazerScience2018}: trustworthiness values can be used by social platforms to limit reader exposure to content from untrustworthy sources\footnote{https://transparency.fb.com/it-it/policies/improving/timeline/} and, conversely, highlight credible sources of information~\cite{Nadarevic2020}. 
Since the procedure of assessing a publisher's trustworthiness is time-consuming and traditionally requires experienced annotators~\cite{doi:10.1073/pnas.1806781116}, in this paper, we tried to approach the problem through the automatic classification of articles. 
Similar work to ours is in~\cite{bohavcek2022fine}, where articles are annotated by hand (while we rely on the external source NewsGuard), and Bert is used for classification, amongst other language models. Like our work, this also considers multiclass classification at the article level. The articles were indeed associated with 4 trustworthiness intervals. Unfortunately, a fair comparison between our results and their performance is impossible because the articles in~\cite{bohavcek2022fine} are in Czech (the best classification performance in F1 is 0.52). The work in~\cite{DBLP:conf/aaai/Przybyla20} considers a labelled news dataset in much the same way as ours: the tag of the individual news item is inherited from the trustworthiness of its source. Specifically, untrustworthy sources are collected from Politifact\footnote{https://www.politifact.com} while trustworthy sources are extracted from a study conducted by the Pew Research Center in~\cite{Mitchell2014}.
The classification task uses multiple models, including BERT, and the accuracy achieved in testing is $0.99$ for articles whose source has already been known by the model under training. The strong difference between~\cite{DBLP:conf/aaai/Przybyla20} and the current work lies in the fact that in the former, the classification is binary, trustworthy vs. untrustworthy, whereas we consider a multiclass classification task.

On a very inspiring final note, a recent paper tested Chat-GPT's API to rank the reputability of an online news publisher, and the model obtains an Area Under the Curve of 0.89 in a binary ranking scenario (trusted/not trusted)~\cite{MenczerGPT2023}. Of course, the judgments are based on sources known to the model during training, but the first results suggest a LLM can also help analysts with the task of ranking the reputability of an online media outlet.

\section{Conclusions}
This paper examines the quality of the online news landscape, with particular reference to the level of trustworthiness of the news source. 
Many organizations, often formed by journalists and communication experts, have been trying for years to guide readers to read online media more trustworthyly by assigning trustworthiness ratings to various online newspapers. Of course, this process requires experienced annotators and is time-consuming. In this paper, we have tried to speed up the work of these organizations by evaluating the quality of an automatic ranking of an article's trustworthiness. The results are very promising when compared to the few existing related works.

Our approach is not intended to replace the careful procedures of journalistic organizations that invest much time and manpower in ranking online news media. Instead, as introduced at the beginning of the paper, we believe that the proposed article ranking can provide such organizations with initial guidance, both in selecting articles for human annotators to analyze and gaining insight into completely unfamiliar media outlets. In addition, our model can suggest to users the similarity - or otherwise - of the news they are reading to news from less reputable sources. 

Our work contributes to real-world applications to combat the spread and impact of low-credibility content.
However, the proposed work can be extended in three main directions: i) enriching the dataset with more sources, including non-English ones, and adding more articles per source; ii) exploring models other than BERT; and iii) integrating eXplainable Artificial Intelligence (XAI) techniques to understand textual differences between articles with different levels of trustworthiness and whether there is a specific reason why some sources are not classified correctly.

\begin{acks}
\small
Work partially supported by project SERICS (PE00000014) under the NRRP MUR program funded by the EU - NGEU; by the Integrated Activity Project TOFFEe (TOols for Fighting FakEs) \url{https://toffee.imtlucca.it/}; by the IIT-CNR funded Project re-DESIRE (DissEmination of ScIentific REsults 2.0).
\end{acks}

\bibliographystyle{ACM-Reference-Format}
\bibliography{bibliography} 


\section{Appendix}
The trustworthiness classification model is available  \href{https://sites.google.com/imtlucca.it/trust-online-news-publishers/home}{here}.

\begin{figure}
    \includegraphics[width=0.45\textwidth]{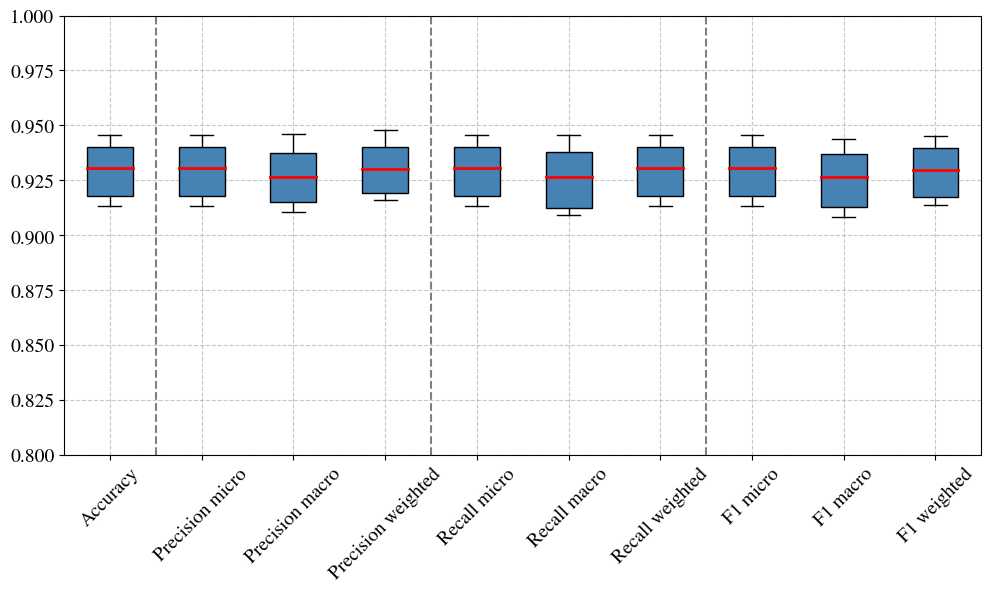}
    \caption{Complete evaluation results for Topic Detection}\label{fig:topicScores}
\end{figure}

\begin{figure}[H]
    \includegraphics[width=0.45\textwidth]{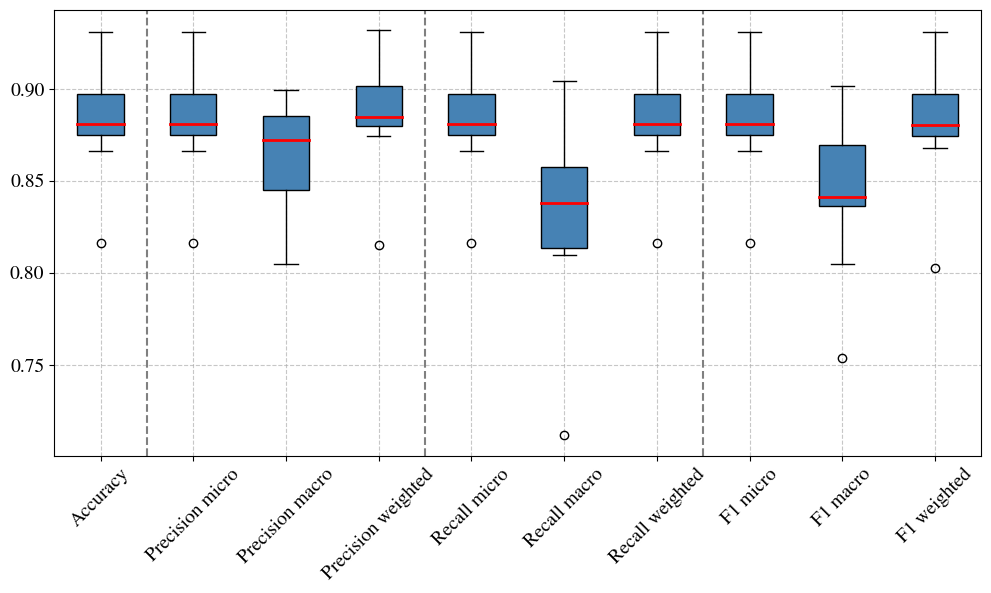}
    \caption{Complete evaluation results for Trustworthiness Detection}\label{fig:trustScores}
\end{figure}

For the convenience of the reader, Table~\ref{tab:domains} lists the domains of the news outlets used to create our dataset. Topic and source trustworthiness tags cannot be released because they are proprietary to NewsGuard and licensed to the paper's authors. We would be happy to agree with NewsGuard to release some of the tags. Figure~\ref{fig:topicScores} and Figure~\ref{fig:trustScores} show the performances obtained for the Topic Classification Task and the Trustworthiness Classification Task regarding Accuracy, Precision, Recall and F1.

\begin{table}
    \centering
    \caption{\label{tab:domains}List of news domains considered in the final dataset}
    \begin{tabular}{>{\fontsize{6}{10}\selectfont}l>{\fontsize{6}{10}\selectfont}l>{\fontsize{6}{10}\selectfont}l}
        \hline
        \textbf{Domain names} & & \\ \hline
        12up.com & aclu.org & americanprogress.org \\
        bizpacreview.com & californiahealthline.org & cbssports.com \\
        celebritiesdeaths.com & clutchpoints.com & nih.gov \\
        news-front.info & ewg.org & famadillo.com \\
        flagandcross.com & harmonyhustle.com & historyfact.in \\
        labourlist.org & nationalrighttolifenews.org & now8news.com \\
        nowtheendbegins.com & on3.com & outsideonline.com \\
        politichome.com & powerofpositivity.com & psypost.org \\
        pulsetoday.co.uk & realrawnews.com & sbnation.com \\
        scoopearth.com & skepticalraptor.com & sportscasting.com \\
        theamericanmirror.com & thecovidblog.com & thegrayzone.com \\
        thelibertytimes.com & thepatriotjournal.com & theplayerstribune.com \\
        theringer.com & thetentacle.com & tnewsnetwork.com \\
        trendingpoliticsnews.com & truthdig.com & unz.com \\
        wavefunction.info & consciousreminder.com & countylocalnews.com \\
        deadspin.com & dreddymd.com & drjockers.com \\ \hline
    \end{tabular}
\end{table}

\end{document}